\documentclass[%
 aip,
  pop, % Physics of Plasmas substyle of AIP, no titles in citations
 amsmath,amssymb,
 preprint,%
]{revtex4-1}
\usepackage{CJK}

\usepackage{ulem}
\usepackage[T1]{fontenc}
\usepackage{calligra}

\usepackage{graphicx}% Include figure files
\usepackage{dcolumn}% Align table columns on decimal point
\usepackage{bm}% bold math

\usepackage{times}
\usepackage{verbatim}
\usepackage{graphicx}
\usepackage{graphics,epsfig}

\usepackage{theorem}
\usepackage{makeidx}
\usepackage{amsmath}
\usepackage{epic}
\usepackage{amscd}

\usepackage{bbm}
\usepackage{xy}
\usepackage{amssymb}
\usepackage{epsfig}
\newcommand{\ignore}[1]{}  % a command which does nothing.
\def \phi {\mbox{$\varphi$}}

\newsavebox{\astrutbox}
\sbox{\astrutbox}{\rule[-5pt]{0pt}{20pt}}

\newcommand{\bea}{\begin{eqnarray}}
\newcommand{\eea}{\end{eqnarray}}
\newcommand{\beq}{\begin{equation}}
\newcommand{\eeq}{\end{equation}}

\begin{document}
%\begin{CJK*}{GB}{gbsn}

\preprint{AIP/123-QED}

%\title[manuscript for PoF]{Statistical mechanics of truncated gyrokinetic collisionless plasma}% Force line breaks with \\

\title[Chiral ensembles of statistical hydrodynamics: ``order function'' for turbulence transfer scenarios]{Note on specific chiral ensembles of statistical hydrodynamics: ``order function'' for transition of turbulence transfer scenarios}

% \thanks{Footnote to title of article.}

\author{Jian-Zhou Zhu}
\affiliation{%\mbox{
Su-Cheng Centre for Fundamental and Interdisciplinary Sciences, Gaochun, 211316 Jiangsu, China and Li Xue Center, Gui-Lin Tang Lab., 47 Bayi Cun, Yong'an, 366025 Fujian, China}%}

%\date{Received 23 June 2010 and accepted 19 October 2010 by {\it Physics of Plasmas}}
%\date{November 4, 2010, to be published in {\it Physics of Plasmas}}
\date{\today}% It is always \today, today,

\begin{abstract}
Hydrodynamic helicity signatures the parity symmetry breaking, chirality, of the flow. Statistical hydrodynamics thus respect chirality, as symmetry breaking and restoration are key to their fundamentals, such as the spectral transfer direction and its mechanism.
Homochiral sub-system of three-dimensional (3D) Navier-Stokes isotropic turbulence has been numerically realized with helical representation technique to present inverse energy cascade [Biferale et al., Phys. Rev. Lett., {\bf 108}, 164501 (2012)]. The situation is analogous to 2D turbulence where inverse energy cascade, or more generally energy-enstrophy dual cascade scenario, was argued with the help of a negative temperature state of the absolute equilibrium by Kraichnan. Indeed, if the helicity in such a system is taken to be positive without loss of generality, a corresponding negative temperature state can be identified [Zhu et al., J. Fluid Mech., {\bf 739}, 479 (2014)].
%The technique used in these works is helical representation to enable one to study the isolated or restricted ensemble of statistical hydrodynamics composed of pure helical modes (``chiroids'').
%These findings might lead to naive expectation that two sign-definite invariants is the sufficient or necessary condition for the dual-cascade scenario.
Here, for some specific chiral ensembles of turbulence, we show with the corresponding absolute equilibria that even if the helicity distribution over wavenumbers is sign definite, different \textit{ansatzes} of the shape function, defined by the ratio between the specific helicity and energy spectra $s(k)=H(k)/E(k)$, imply distinct transfer directions, and we could have inverse-helicity and forward-energy dual transfers (with, say, $s(k)\propto k^{-2}$ resulting in absolute equilibrium modal spectral density of energy $U(k)=\frac{1}{\alpha +\beta k^{-2}}$, exactly the enstrophy one of two-dimensional Euler by Kraichan), simultaneous forward transfers (with $s(k)=constant$), or even no simply-directed transfer (with, say, non-monotonic $s(k) \propto \sin^2k$), besides the inverse-energy and forward-helicity dual transfers (with, say, $s(k)=k$ as in the homochiral case).
%All such cases satisfy the natural condition $s(k)=H(k)/E(k)\le k$ and may correspond to some spontaneous physical states under specific time regime (say, metastable state) with some specific physical constraints. Should such specific ensembles, from the (turbulent) dynamics constrained by the shape function, be numerically realized, corresponding transfer scenarios will be observed.
%Such subensemble may also be meaningful in the consideration of metastable states \cite{etcLebowitz}; different ensembles other than the final ensembles or subensembles of the final ones may also be meaningful for metastable or long-lived transient states, such as the (quasi-)2D dynamics of rotating fluid \cite{YKR02,Bourouiba08}.

%
%Valid PACS numbers may be entered using the \verb+\pacs{#1}+ command.
\end{abstract}

% \pacs{Valid PACS appear here}% PACS, the Physics and Astronomy
% \keywords{Suggested keywords}%Use showkeys class option if keyword

\maketitle
%\end{CJK*}

\thispagestyle{plain}

\section{\label{sec:Introduction}Introduction}
Statistical hydrodynamics respect chirality, the absence of parity symmetry signatured by helicity, as symmetry breaking and restoration are key to its fundamentals (c.f., the monograph by Frisch \cite{FrischBook} which starts with the introductory discussion of ``turbulence and symmetry''.) A fundamental issue of turbulence is the spectral transfer direction and its mechanism. One remarkable insight by Kraichnan \cite{k67,k73} is that, for two-dimensional (2D) turbulence constrained by the energy and enstrophy, a negative-temperature state with condensation of energy at large scales provides an inverse energy cascade mechanism. But, for the normal 3D isotropic helical turbulence, due to the fact that helicity distribution (in physical space or wavevector space) is not sign definite, there is no energy condensation state to imply dual cascades analogous to that of 2D turbulence. Nevertheless, when the dynamics are restricted to pure helical modes of sign-definite helicity the analogy would be a close one \textit{per se} \cite{w92,bmt13}, and indeed a corresponding energy-condensation \textit{absolute equilibrium} (AE) state can be identified as shown by us in Ref. \onlinecite{zyz14} where, however, we also specified a restriction without sign-definite helicity but with an energy condensation state which might also imply the possibility of an inverse cascade of energy in turbulence simulations with the corresponding restriction. However, one might still speculate that any restriction, for whatever physical or artificial constraint, with sign-definiteness of helicity would be sufficient to have a condensation implying inverse cascade of energy. Such a speculation is neither valid as will be explicitly shown with several specific chiral hydrodynamic ensembles.
%, and in this case one might simply apply to the notion of restricted ensemble identified with a specific region in the phase space or imagine that by some appropriate numerical scheme the specified ensemble itself can be generated.

%\subsection{Helical representation and the absolute equilibria}
%{\it Specific chiral Gibbs ensembles}---
A convenient tool to look into the issue of hydrodynamic parity symmetry breaking, chirality, is the helical decomposition.
Morses (1971) has used the complete and orthogonal eigenfunctions of the curl operator for expansion of a vector function, a field not necessarily divergence free/incompressible. And ``the original Helmholtz theorem has been sharpened in two ways. First, two irrotational vectors have been introduced in the decomposition of a general vector, each of which is the curl of its own vector potential. Second, a procedure has been given for obtaining the vector and scalar potentials''. Here the ``Helmholtz theorem'' is what decomposes a field into the rotational/transverse and irrotation/longitudinal parts, representable respectively by the curl of a potential vector and the gradient of a scalar potential. The longitudinal component of the field is represented with the eigenfunctions corresponding to the eigenvalue $0$ and is uncoupled from the transverse part.

Then, for a 3D transverse vector field $\bm{v}$ the helical mode/wave representation in Fourier space reads \cite{Moses71,JonLee79,CambonJacquinJFM89,w92}
\begin{equation}\label{eq:FourierHelical}
\bm{v}%=\sum_{\bm{k}} \hat{\bm{v}}(\bm{k})e^{\hat{i}\bm{k}\cdot \bm{r}}
=\sum_c \bm{v}^c=\sum_{\bm{k},c} \hat{\bm{v}}^c(\bm{k}) e^{\hat{i}\bm{k}\cdot \bm{r}}=\sum_{\bm{k},c} \hat{v}^c(\bm{k})\hat{\bm{h}}_c(\bm{k})e^{\hat{i}\bm{k}\cdot \bm{r}}.
\end{equation}
Here $\hat{i}^2=-1$ and $c^2=1$ for the chirality indexes $c=$ ``+'' or ``-''. [For consistency of notation, every complex variable wears a hat and its complex conjugate is indexed by ``*''.] For convenience, we normalize the box to be of $2\pi$ period and that $k\ge 1$. The helical mode bases (complex eigenvectors of the curl operator) have the following properties $$\hat{i}\bm{k}\times \hat{\bm{h}}_c(\bm{k})=ck\hat{\bm{h}}_c(\bm{k}),$$ $$\hat{\bm{h}}_c(\bm{-k})=\hat{\bm{h}}_c^*(\bm{k})=\hat{\bm{h}}_{-c}(\bm{k})$$ and $\hat{\bm{h}}_{c_1}(\bm{k})\cdot\hat{\bm{h}}_{c_2}^*(\bm{k})=\delta_{c_1,c_2}$ (Euclidean norm). $\hat{\bm{h}}_c(\bm{k})e^{\hat{i}\bm{k}\cdot \bm{r}}$ is the eigenfunction of the curl operator corresponding to the eigenvalue $ck$. Or, with the case $c=0$ also included for the compressible field, the variable $c$ ``itself may be considered to be the eigenvalue of the operator $(-\nabla^2)^{-1/2}\nabla\times$ when this operator is properly interpreted.''\cite{Moses71} The bases can be simply constructed as \cite{Greenspan90,w92} $$\hat{\bm{h}}_c(\bm{k})=(c\hat{i}\bm{p}+\bm{p}\times\bm{k}/k)/(\sqrt{2}p),$$ with $\bm{p}$ being perpendicular to $\bm{k}$. The structure $\hat{\bm{h}}_c(\bm{k})e^{\hat{i}\bm{k}\cdot\bm{r}}$ is common in inertial waves of rotating fluids and cyclotron waves of plasmas, being circularly
polarized, with $c=\pm$ representing opposite chirality. For better or alternative physical intuition, we may conveniently call $$\breve{\bm{v}}^c(\bm{r}|\bm{k})=\hat{v}^c(\bm{k})\hat{\bm{h}}_c(\bm{k})e^{\hat{i}\bm{k}\cdot \bm{r}}+c.c.,$$ with $c.c.$ for ``complex conjugate'', a ``chiroid'' which is maximally/purely helical or of highest degree of chriality, since the helicity contribution of it is $$\nabla\times\breve{\bm{v}}^c(\bm{r}|\bm{k})\cdot\breve{\bm{v}}^c(\bm{r}|\bm{k})=2ck|\hat{v}^c(\bm{k})|^2=ck|\breve{\bm{v}}^c(\bm{r}|\bm{k})|^2;$$ other corresponding chemistry terminologies, such as enantiomer, enantiopure and racemic etc. may also be tentatively borrowed for this purpose.

The ideal statistical hydrodynamics velocity field $\bm{u}$ is governed by
\begin{equation}\label{eq:rtr}
    \partial_t \bm{u}=\bm{u}\times (\nabla\times\bm{u})-\nabla P; \ \nabla \cdot \bm{u}=0,
\end{equation}
where $P$ is the pressure. This equation reads in the Fourier space with helical representation
\cite{Moses71,JonLee79,w92}
\begin{eqnarray}
    \partial_t \hat{u}_{c_{\bm{k}}}&&=
    \sum_{\bm{k}=\bm{p}+\bm{q}}\sum_{c_{\bm{p}},c_{\bm{q}}}\hat{C}_{\bm{k}\bm{p}\bm{q}}^{c_{\bm{k}}c_{\bm{p}}c_{\bm{q}}} \hat{u}_{c_{\bm{p}}}\hat{u}_{c_{\bm{q}}} \ \text{with}  \   \frac{\hat{C}_{\bm{k}\bm{p}\bm{q}}^{c_{\bm{k}}c_{\bm{p}}c_{\bm{q}}}}{c_{\bm{q}}q-c_{\bm{p}}p}= \frac{\hat{\bm{h}}_{c_{\bm{p}}}\times \hat{\bm{h}}_{c_{\bm{q}}} \cdot \hat{\bm{h}}^*_{c_{\bm{k}}}}{2}.  \label{eq:uk}
\end{eqnarray}
[For notational convenience, here and sometimes later, we denote the chirality coming with $\bm{k}$ with $c_{\bm{k}}$ and thus $\hat{\bm{h}}_{c_{\bm{k}}}$, and similarly for those with $\bm{p}$ and $\bm{q}$.] The symmetry in the above coupling coefficient $\hat{C}_{\bm{k}\bm{p}\bm{q}}^{c_{\bm{k}}c_{\bm{p}}c_{\bm{q}}}$ formally leads to the detailed conservation laws of energy and helicity among each triad $$\{[\pm\bm{k},c_{\bm{k}}];[\pm\bm{p},c_{\bm{p}}];[\pm\bm{q},c_{\bm{q}}]\} \ \text{with} \ \bm{k}+\bm{p}+\bm{q}=0.$$
The \textit{mean} energy $\mathcal{E}=\tilde{\sum}_{\bm{k}}U(k)=\tilde{\sum}_{c,\bm{k}}U^c(k)$ and helicity $\mathcal{H}=\tilde{\sum}_{\bm{k}}Q(k)=\tilde{\sum}_{\bm{k},c}Q^c(k)$, with the tilde `$\tilde{\bullet}$' denotes the Galerkin truncation, i.e., limiting to a subset of the wavenumber set,  come from
\begin{equation}\label{eq:QU}
U^c(k)=cQ^c(k)/k=\langle|\hat{u}_c(\bm{k})|^2\rangle/2,%\nonumber
\end{equation}
where $\langle \bullet \rangle$ denotes the mean, per unit volume or in the statistical sense (by assuming ergodicity, say). 
Here we have used $k$ to replace $\bm{k}$ in the isotropic (modal) spectral densities $U^{(c)}$ and $Q^{(c)}$ which can be conveniently obtained from the Gibbs ensemble, with the distribution $\propto \exp\{-(\alpha \mathcal{E}+\beta\mathcal{H})\}$:
\begin{eqnarray}
 U^c(k) %&&
 =1/(\alpha+c\beta k), \ %\label{eq:UK}\\
   Q^c(k) %&&
   %=\frac{c k}{\alpha+c \beta k}
   = c k U^c(k). \label{eq:QK}
\end{eqnarray}
Several remarkable findings about one-chiral-sector-dominated state (OCSDS) have been pointed out \cite{zyz14}:
\begin{enumerate}
  \item In the chirally symmetrical truncation case, the pole $k_p=\alpha/|\beta|$, say, in the $c=+$ chiral sector with $\beta<0$, indicates that energy and helicity subject to dissipation are more persistent in this sector which was observed by Chen et al. \cite{CCE03}, as a kind of ``implicit'' or ``second order'' OCSDS;
  \item when $k_p$ is restricted to small-$k$ regime, say, close to the lower bound of the truncation wavenumber $k_{min}$, {\it i.e.}, all the chiroids with $k_p<k<k_{max}$ in this sector is further truncated, large-scale energy condensation is carried by the left alien(s) at $k_p$ and of course is OCSDS;
  \item when one chiral sector is thoroughly truncated, of course an extremal case of OCSDS, energy condensation at largest scales is possible for the surviving sector $c$.
\end{enumerate}

\section{Specific chiral ensembles}
All the above cases are specified by the explicit schemes of the truncation on the chiroids, and, with $s(k)=ck$, the last homochiral case has
\begin{eqnarray}
 U(k)=U^c(k)=Q(k)/s(k)
 =\frac{1}{\alpha+\beta s(k)}. \label{eq:QU}
\end{eqnarray}
Here, the \textit{shape function} $s(k)$ defines the relation between the spectra of the two rugged invariants, and it is for the ensemble specific for or restricted to the homochiral chiroids.

In the above we have indicated the generalization of the specific ensembles with
\begin{equation}\label{eq:squ}
%\hat{\bm{\omega}}(\bm{k})\cdot \hat{\bm{u}}^*(\bm{k})=
Q(\bm{k})=s(k)U(\bm{k})
\end{equation}
which can be realized by
\begin{equation}\label{eq:su}
%\hat{\bm{\omega}}(\bm{k})\cdot \hat{\bm{u}}^*(\bm{k})=
\hat{i}\bm{k}\times\hat{\bm{u}}(\bm{k})\cdot \hat{\bm{u}}^*(\bm{k})=s(k)|\hat{\bm{u}}(\bm{k})|^2.
\end{equation}
The freedom of $s(k)$ comes from the indefiniteness of the relation and the cancelation between the velocities of opposite chiralities in Eq. (\ref{eq:uk}). It depends on the specific physical constraint or numerical scheme.
Note that we \textit{assume} that the above relation can be realized by some restriction on the triadic interactions in Eq. (\ref{eq:uk}) and consequently the conservation laws (due to the detailed ones\cite{k73}) and Liouville theorem are inherited. [One particular case with $s(k)=\pm k$ is obviously realizable as already shown \cite{w92,bmt13}, while schemes for more general ones remain to be discovered.]
We then have $$\alpha |\hat{\bm{u}}(\bm{k})|^2 +\beta \hat{i}\bm{k}\times\hat{\bm{u}}(\bm{k})\cdot \hat{\bm{u}}^*(\bm{k})=[\alpha+\beta s(k)]|\hat{\bm{u}}(\bm{k})|^2$$ in the Gibbs ensemble, thus deriving Eq. (\ref{eq:QU}), $U(k)=Q(k)/s(k)=\frac{1}{\alpha+\beta s(k)}$. Using $\hat{\bm{u}}=\hat{u}_+\hat{\bm{h}}_+ + \hat{u}_-\hat{\bm{h}}_-$, we have $[k-s(k)]U^+=[k-s(k)]U^-$; and further explicit expressions of $U^c$ with $U=U^++U^-$.
%and that
%$$Q(k)/s(k)=U(k)=\frac{1}{\alpha+\beta s(k)}.$$
%Due to $Q^c(k)=ckU^c(k)$ in $Q(k)=\sum_cQ^c(k)$ and $U(k)=\sum_cU^c(k)$, equivalently $$U^+(k)=\frac{k+s(k)}{k-s(k)}U^-(k)=\frac{k+s(k)}{2k} \frac{1}{\alpha+\beta s(k)}.$$
Since we have not explicitly proposed a scheme to select specific interaction from Eq. (\ref{eq:uk}) nor shown that a restricted ensemble, say, defined by a selected region in the whole phase space,\cite{etcLebowitz}, satisfies the dynamics of Eq. (\ref{eq:uk}), especially the conservation laws and Liouville theorem, it makes sense to make it more concrete by considering a statistical model as follows. The two chiral sectors of such a field may be connected with a random multiplier $M$: \begin{equation}\label{eq:M}
\hat{u}_+(\bm{k})\eqcirc M(\bm{k})\hat{u}_-(\bm{k})
\end{equation}
where $\eqcirc$ indicates the same statistical properties, that is, ``equality in law'', as is widely used to describe turbulence cascade between two scales \cite{FrischBook}.
Taking $M(k)$ independent of $\hat{u}_-(k)$ and using $\hat{\bm{u}}=\hat{u}_+\hat{\bm{h}}_+ + \hat{u}_-\hat{\bm{h}}_-$ in Eq. (\ref{eq:su}), we have
\begin{equation}\label{eq:M2}
\frac{k+s(k)}{k-s(k)}=\langle |M(k)|^2\rangle_M,%;
\end{equation}
%the restricted Gibbs ensemble immediately gives $$U^-(k)=\langle\frac{1}{\alpha [M^2(k)+1] + \beta k[M^2(k)-1]}\rangle_M,$$
where $\langle \cdot \rangle_M$ means the average over the disorder of $M$.
Such a multiplier helpfully indicates at least two things about the internal structures of the ensembles: One is that the shape function $s(k)$ may come from a statistical-average effect, that is, \textit{Eq. (\ref{eq:su}) may be of a more general form with the fixed $s(k)$ replaced by a stochastic one}; the other is that it may lead to a practical numerical scheme for preparing the ensemble, at least \textit{for possible phenomenological studies of chiral turbulence}. If we assume Eq. (\ref{eq:M}) comes from the original dynamics, Eq. (\ref{eq:uk}), it would then be very easy to construct the specific ensembles by simply specifying the statistics of the multiplier $M$, just as in the phenomenological random cascade models.\cite{FrischBook}

Note in particular that even for the non-homochiral ``fundamental triad-interaction system'' \cite{JonLee79} with a single interacting triad on the right hand side of Eq. (\ref{eq:uk}), though $Q^{c_{\bm{k}}}(\bm{k})=c_{\bm{k}}kU^{c_{\bm{k}}}(\bm{k})$ for each unichiral/enantiopure $\bm{k}$, $c_{\bm{k}}k$ is not necessarily monotonic with different signs for different $\bm{k}$. \cite{footnoteHUc}

From now on we focus on the sign-definite, actually, without loss of generality, the positive-definite-$Q$ case. That is, $s(k)>0$. The homochiral case corresponds to $s(k)=k$, allowing a negative-temperature energy condensation state \cite{zyz14}, which implies inverse-energy and forward-helicity cascades of turbulence. Assuming local cascade and the applicability of simple dimensional analysis, we find the inertial range scaling law $Q(k)\propto k^{-10/3}$ for the forward helicity cascade and $U(k)\propto k^{-11/3}$, i.e., the familiar Komogorov one-dimensional energy spectrum $E(k)\propto k^{-5/3}$, for the inverse energy cascade \cite{BrissaudETC73,w92}, and one numerical realization is to simply truncate one of the chiral sectors thoroughly \cite{bmt13}. The reason for the existence of a condensation state of the absolute equilibrium is further elaborated as follows: From Eq. (\ref{eq:QU}) with $s(k)=k$, when the untruncated $\bm{k}$ satisfies $k_{min}\le |\bm{k}|\le k_{max}$, $U(k)>0$ leads to $\alpha+\beta k>0$; for $\alpha<0$, $k>\frac{-\alpha}{\beta}$. So, the pole $k_p=-\alpha/\beta$ can approach $k_{min}$ from below and that a smallest ratio between helicity and energy $\mathcal{H}/\mathcal{E}=k_{min}$ with energy concentrating at $k_{min}$ is approached. Similarly, for other monotonically increasing shape function such as $s(k)=k^n$ with $1>n>0$ (note that, as assumed, here and below $k>1$; or, we restrict ourselves to the scales smaller than the one used for normalization), we can reach the same conclusion with $k_p=(\frac{-\alpha}{\beta})^{1/n}$.

Whenever there is cancelation between positive and negative helicities, $s(k) < k$. As we have seen from the above, the monotonicity of the shape function $s(k)$ is crucial for the implication of turbulence transfers from the absolute equilibrium. While $s(k)=k$ or others are monotonically increasing, the natural condition $s(k)\le k$ leaves a large space for other monotonically decreasing or non-monotonic shape functions, which is the main point we want to emphasize in the following.
%\begin{enumerate}
%  \item
  For $$s(k)=k^{-2},$$ we have $$U(k)=\frac{1}{\alpha +\beta k^{-2}},$$ exactly the form of the absolute-equilibrium enstrophy spectrum of 2D Euler obtained by Kraichnan \cite{k67}! In words, when $s(k)\propto k^{-2}$, the relation is in exact analogy to 2D turbulence, with $U(k)$ playing the role of (modal) enstrophy spectrum density and $Q(k)$ the energy spectrum density there. Now, since $s(k)=k^{-2}$ is monotonically decreasing, $U(k)>0$ gives $k>k_p=(\frac{\alpha}{-\beta})^{-1/2}$ for $\beta<0$. Again, $k_p$ can approach $k_{min}$ from below to reach a smallest $\mathcal{E}/\mathcal{H}=k_{min}^2$ with helicity concentrating at $k_{min}$, which immediately implies inverse-helicity and forward-energy cascades for such restricted ensemble of turbulence, \textit{i.e.}, for a turbulence with particular physical constraint leading to such a specified spectral relation. $s(k)=k^{-2}$ indicates that helicity is relatively emphasizing large scales and energy small ones, so the normal viscosity or hyperviscosity emphasizing small scales would effectively dissipate energy; hypoviscosity emphasizing large scales will then effectively remove helicity there. The hypoviscosity or simply the helicity condensation at largest scales will then lead to the transfer/cascade of helicity injected into the intermediate or small scales. Assuming local cascade and the applicability of simple dimensional analysis, we find the inertial range scaling law $U(k)\propto k^{-11/3}$, i.e., the familiar Kolmogorov one dimensional energy spectrum $E(k)\propto k^{-5/3}$, for the forward energy cascade and $Q(k)\propto k^{-10/3}$ for the inverse helicity cascade.
%  \item

  It is also easily seen that when $s(k)=const.$, both energy and helicity are equipartitioned over each Fourier mode. For turbulence state they are expected to simultaneously cascade forwardly to small scales where dissipation happens \cite{k67,k73}. Dimensional analysis for the local cascade predicts a scaling law $k^{-11/3}$ for both $U(k)$ and $Q(k)$ \cite{BrissaudETC73,k73}.
  %This is typically observed in the most ``natural'' simulations of helical turbulence with both helicity and energy injected at large scales of the full Navier-Stokes.
%  \item

  From the above three cases, we see immediately that when the shape function is not monotonic, say, $$s(k)\propto \sin^2k,$$ the argument of the tendency of relaxation to absolute equilibrium can not determine a simply-directed transfer for turbulence. An exact sinusoidal shape function might appear to be somewhat strange, but it just represents a possible situation that the dynamics at different scales are subjected to very different constraints, leading to distinct degrees of cancelations between the positive and negative sectors, so that $s(k)$ may first increase (say, as $k$ as in a previous case), then decrease (say, as $k^{-2}$ as in a previous case), and then increase,..., with $k$.
%\end{enumerate}

\section{Concluding remarks}
%\textit{
Note that we are not saying that in a simulation of turbulence, say, the conventional isotropic turbulence, one can see different cascades for any simply selected samples of fluctuations bearing the above $s(k)$; but, should one be able to numerically realize such a restricted ensemble from the dynamics constrained by the shape function \cite{footnoteLJ} with the restriction be dynamically isolated, in the sense that the detailed conservation laws and Liouville theorem are preserved through the original triadic interactions, to justify the identification of a dynamical ensemble, the corresponding spectral transfer scenarios should be observed.
%}
Such specific (sub)ensembles, or different ensembles other than the final ensembles or subensembles of the final ones, may also be meaningful in the consideration of metastable states \cite{etcLebowitz} or for long-lived transient states, such as the (quasi-)2D dynamics of rotating fluids. Indeed, Yamazaki et al. \cite{YKR02} found that the relaxation to the final full Euler absolute equilibrium is delayed by rotation, and Bourouiba \cite{Bourouiba08} clearly shows that under rapid rotation absolute equilibria of the decoupled systems with extra constraints of conservation laws (but with helicity omitted), predicted by the resonant wave theory, can be identified during the long-lived transient stage before the threshold time $t_*$ (after which the resonant wave theory breaks down). Following these, as will be reported in another communication, a helical absolute-equilibrium ensemble of the slow modes can be shown to support the somewhat novel inverse energy transfer (to large scales) of vertically averaged vertical velocity in the turbulence simulation by Chen et al. \cite{CCEH05}. Numerical analyses in the 1970s by Orszag and Lee et al. (see Ref. \onlinecite{JonLee79} and references therein) reveal that subsystems composed of the fundamental triadic interactions present rich ``surprising'' dynamical properties, such as those relevant to extraneous or quasi-constants and to ergodicity or mixing properties, which indicates that the specific chiral ensembles and corresponding turbulence behaviors should be possible to be dynamically realized either by specific artificial restriction or by natural emergence of the system under specific physical conditions (such as rotation or stratification etc.)

In conclusion, we can try to introduce a specific ``order parameter'' $\epsilon$ in the above shape functions, such as $$s(k)=\epsilon k, \ s(k)=\epsilon k^{-2} \ \text{and} \ s(k)= \epsilon \sin^2 k,$$ without changing the results about the implication of cascade scenario, which indicates that no simple ``phase transition'' can be concluded according to the variation of a single ``order parameter''; what we need is an ``order function'', the specific shape function or its normalization as a measure of relative helicity $s(k)/k$, to describe the symmetry breaking and ``phase transition'' of the cascade scenario. All these cases show parity symmetry breaking, i.e., chirality, and the helicity is sign definite at all scales (with $\epsilon>0$, say.) Thus, it appears that a recent paper by Herbert,\cite{cherbert} which only narrowly extends our previous result of the negative-temperature homochiral Euler absolute equilibrium is incomplete; especially, one should, as shown here, avoid the incorrect conclusion ``that when helicity is sign definite at all scales, an inverse cascade is expected for the energy'' and, as already suggested in Zhu, Yang and Zhu (summarized as the second item in the end of the introductory discussion of this paper),\cite{zyz14} should be careful with the statement that ``W[w]hen sign-definiteness is lost, even for a small set of modes, this [inverse] cascade disappears and there is a sharp phase transition to the standard helical equipartition spectra.'' Those inappropriate theoretical conjectures are due to the simple logical pathology by taking, just as we have warned in the introductory discussion, a very special sufficient condition for condensation of the absolute equilibrium, from the particular case with $s(k)=\epsilon k$ among several others as listed here, as the necessary condition, without considering other statistical ensembles. Finally we remark that although such statistical ensembles and the corresponding turbulent flows look artificial and remain a challenge to connect with realities, their theoretical value is obvious. Such possibilities of various specific or restricted chiral turbulence ensembles with different transfer scenarios should not be surprising at all, since the dynamics, Eq. (\ref{eq:uk}), contain triad interactions capable of such scenarios.\cite{w92} Should such specific ensembles, from the (turbulent) dynamics constrained by the shape function, be numerically realized, corresponding transfer scenarios will be observed. Further theoretical and numerical studies of the relevant absolute equilibria and the corresponding turbulent flows for more details, in which possible devil could hide, following Lee \cite{JonLee79}, Cichowlas et al. \cite{CichowlasPRL05}, Bos and Bertoglio \cite{bb06} (extending analytical models such as that of Cambon and Jacquin \cite{CambonJacquinJFM89} to restricted ensembles) and Biferale et al. \cite{bmt13} are promising. Especially, though a random multiplier model is suggested, we have not explicitly given the general schemes to construct the specific or restricted ensembles systematically from the original equation (\ref{eq:uk}), which remains open.\\

\section*{Acknowledgement}
This work is partially supported by National Natural Science Foundation of China under Grant No. 11375190. The author thanks for the helpful comments (including the suggestion of citing Ref. \onlinecite{cherbert}) from the anonymous referees. He also acknowledges the interesting conversations with Mr. Xiaoxiao Ma in Ninggao Hi-Tech Park.

%\end{comment}


\begin{thebibliography}{10}

\bibitem%[Frisch(1995)]
{FrischBook}
U. Frisch, Turbulence: The Legacy of Kolmogorov.
Cambridge University Press (1995).

\bibitem%[Kraichnan(1967)]
{k67}
R.~H. Kraichnan, %{\em
  Phys. Fluids %\/}
{\bf 102}, 1417 (1967).% Inertial ranges in two-dimensional turbulence. .

\bibitem{k73}
R. H. Kraichnan, J. Fluid Mech. {\bf 59} 745 (1973).%Inertial-range transfer in two- and three-dimensional turbulence.

\bibitem%[Waleffe(1992)]
{w92}
F. Waleffe, %1992 The nature of triad interactions in homogeneous turbulence.
{\em Phys. Fluids A\/} {\bf 4}, 350 (1992).

\bibitem{bmt13}
L. Biferale, S. Musacchio and F. Toschi, J. Fluid Mech. {\bf 730} 309 (2013).

\bibitem{zyz14}
J.-Z. Zhu, W. Yang, G.-Y. Zhu, J. Fluid Mech. {\bf 739} 479 (2014).

\bibitem%[Moses(1971)]
{Moses71}
H. E. Moses, %1971 Eigenfunctions of the curl operator, rotationally invariant Helmholtz theorem and applications to electromagnetic theory and fluid mechanics. {\em
SIAM ~(Soc. Ind. Appl. Math.) J. Appl. Math. {\bf 21}, 114 (1971).

\bibitem{JonLee79}
J. Lee, Phys. Fluids. {\bf 22} 40 (1979).

\bibitem%[Cambon \& Jacquin(1989)]
{CambonJacquinJFM89}
%{\sc Cambon, C. \& Jacquin, L.} 1989 Spectral approach to non-isotropic turbulence subjected to rotation.
C. Cambon, L. Jacquin,
J. Fluid Mech. 202, 295 (1989).

\bibitem{Greenspan90}
H. P. Greenspan, The theory of rotating fluids. Breukelen Press (1990).

\bibitem{footnoteHUc}
It should have been self-evident that ``homochiral'' is used for stronger condition than ``unichiral'': it means not only unichiral/enantiopure for each $\bm{k}$ but also uniform handness for all $\bm{k}$.

\bibitem%[Chen, Chen \& Eyink(2003)]
{CCE03}
%{\sc Chen, Q. N., Chen, S. Y. \& Eyink, G. L.} 2003 The joint cascade of energy and helicity in three-dimensional turbulence.
Q. N. Chen, S. Y. Chen and G. L. Eyink,
{\em Phys. Fluids\/} {\bf 15} (2), 361 (2003).

%\bibitem[Frisch(1995)]{FrischBook}
%{\sc Frisch, U.} 1995 Turbulence: The Legacy of Kolmogorov.
%Cambridge University Press.

\bibitem%[Brissaud et al.(1973)]
{BrissaudETC73}
%{\sc Brissaud, A., Frisch, U., Leorat, J., Lesieur, M. \& Mazure, M.} 1973 Helicity cascades in isotropic turbulence. {\em Phys. Fluids\/} {\bf 16}, 1366--1367.
A. Brissaud, U. Frisch, J. Leorat, M. Lesieur and M. Mazure, Phys. Fluids {\bf 16}, 1366 (1973).

\bibitem%[Chen et al.(2005)]
{CCEH05}
%{\sc Chen, Q. N., Chen, S. Y., Eyink, G. L. \& Holm, D.} 2005 Resonant interactions in rotating homogeneous three-dimensional turbulence.
Q. N. Chen, S. Y. Chen, G. L. Eyink and D. Holm,
{\em J. Fluid Mech.\/} {\bf 542} (2), 139 (2005).

\bibitem{footnoteLJ}
Successful computations of the Lennard-Jones particle systems (fluid and gas) have been performed with the restricted canonical ensemble Monte Carlo simulation method for some time: See, {\it e.g.}, D. S. Corti and P. G. Debenedetti, Chem. Eng. Sci. {\bf 49}, 2717 (1994), and, N. Chu, G. Jun, and W. H. Marlow, J. Chem. Phys. {\bf 127}, 154505 (2007)

\bibitem{etcLebowitz}
D. Dhar and J. L. Lebowitz, Europhysics Letters {\bf 92} 20008 (2010) and references therein. In such references, typically the relevance to the metastable state are made, and a familiar material is the window glass.

\bibitem%[Bourouiba(2008)]
{Bourouiba08}
%{\sc Bourouiba, L.} 2008 Model of a truncated fast rotating flow at infinite Reynolds number.
L. Bourouiba,
{\em Phys. Fluid\/} {\bf 20}, 07512 (2008).

\bibitem%[Yamazaki, Kaneda \& Rubeinstein(2002)]
{YKR02}
%{\sc Yamazaki, Y., Kaneda, Y. \& Rubeinstein, R.} 2002 Dynamics of inviscid truncated model of rotating turbulence.
Y. Yamazaki, Y. Kaneda and R. Rubeinstein,
J. Phys. Soc. Jpn. {\bf 71}, 81 (2002).

%\bibitem{ZhuRTjfm}
%J.-Z. Zhu, ``On the statistical mechanics of rotating flows'', J. Fluid. Mech. (submitted).

\bibitem{cherbert}
C. Herbert, Phys. Rev. E {\bf 89}, 013010 (2014).

%\bibitem{ZhuETC14}
%J.-Z. Zhu, ``Hall magnetohydrodynamic harmonic-helicon absolute equilibrium'', Proceedings of 14th European Turbulence Conference, 1 - 4 September 2013, Lyon, France; see also the online two-page conference paper: http://etc14.ens-lyon.fr/openconf/modules/request.php?module=oc\_proceedings\&action=view.php\&a=Poster\&id=454

\bibitem{CichowlasPRL05}
C. Cichowlas, P. Bona\"iti, F. Debbasch, and M. E. Brachet, Phys. Rev. Lett. {\bf 95}, 264502 (2005).

\bibitem{bb06}
W. J. T. Bos and J.-P. Bertoglio, Phys. Fluids, {\bf 18} 071701 (2006).

\end{thebibliography}
\end{document}